\documentclass[preprintnumbers,groupedaddress,nofootinbib,aps,prl,10pt,twocolumn]{revtex4-1}
\usepackage[latin1]{inputenc}
\usepackage{graphicx}
\usepackage{color}
\usepackage{multirow}
\usepackage{tabularx}
\usepackage{amsmath,amssymb}
\usepackage{hyperref}
\usepackage{bbold}
\usepackage{url}
\usepackage{slashed}
\usepackage{subfigure}
\usepackage[usenames,dvipsnames]{xcolor}
\newcommand{\be}{\begin{equation}}
\newcommand{\ee}{\end{equation}}
\newcommand{\bea}{\begin{eqnarray}}
\newcommand{\eea}{\end{eqnarray}}

\usepackage[normalem]{ulem}

\begin{document}

\title{Quantum gravity and Standard-Model-like fermions}
 
\author{Astrid Eichhorn}
\email{a.eichhorn@thphys.uni-heidelberg.de}
\affiliation{
 Institut f\"ur Theoretische Physik, Universit\"at Heidelberg, Philosophenweg 16, 69120 Heidelberg, Germany
}
\author{Stefan Lippoldt}
\email{s.lippoldt@thphys.uni-heidelberg.de}
\affiliation{
 Institut f\"ur Theoretische Physik, Universit\"at Heidelberg, Philosophenweg 16, 69120 Heidelberg, Germany
}

\begin{abstract}
We discover that chiral symmetry does not act as an infrared attractor of the renormalization group flow
under the impact of quantum gravity fluctuations.
Thus, observationally viable quantum gravity models must respect chiral symmetry.
In our truncation, asymptotically safe gravity does, as a chiral fixed point exists.
A second non-chiral fixed point with massive fermions provides a template for models with dark matter. 
This fixed point disappears for more than 10 fermions, suggesting that an asymptotically safe ultraviolet
completion for the standard model plus gravity enforces chiral symmetry.
\end{abstract}

\maketitle
\section{Introduction}
An observationally viable model of quantum gravity must be compatible with the existence of matter and all
its low-energy properties. 
This supplies observational tests for quantum gravity,
as particular assumptions about quantum spacetime could be in conflict with low-energy observations. 
The specific fact that we focus on is the existence of light, chiral fermions.
In the Standard Model, chiral symmetry forbids a microscopic fermion mass term $m_{\psi} \bar{\psi} \psi$.
Fermion masses are generated from Yukawa interactions with the Higgs,
and through chiral symmetry breaking in QCD for the quarks. Thus, fermion masses only emerge at scales
far below the Planck scale.
Here, we explore the interplay between quantum gravity and chiral symmetry, finding indications
that chiral symmetry is a nontrivial observational constraint on models of quantum gravity:
The fermion mass remains a Renormalization-Group (RG)-relevant coupling even under the impact
of gravitational fluctuations.
Thus, if chiral symmetry was broken above the Planck scale, it would not be restored automatically
by the RG flow towards low energies, as it would for an irrelevant coupling.
Accordingly, if chiral symmetry is broken in the ultraviolet (UV), the symmetry-violating effects
are expected to generically grow towards low energies, typically leading to large fermion masses.
As a specific illustration, we will focus on asymptotically safe quantum gravity
\cite{Weinberg:1980gg}, before analyzing models of quantum gravity from an effective-field-theory point of view.
\section{Non-minimally coupled fermions in gravity}
We analyze the RG scale dependence of the fermion mass under the impact of quantum-gravity fluctuations.
We focus on its scaling dimension, which is RG relevant according to canonical counting in the free theory.
If quantum fluctuations of spacetime cannot render it irrelevant, then it is expected to grow towards
the infrared (IR).
In this scenario, the appearence of chiral symmetry at low energies would either be impossible or require
severe fine-tuning, unless the microscopic model of quantum gravity contained a mechanism to
impose exact chiral symmetry.
Crucially, the RG flow generates all terms that are compatible with the symmetries.
Thus, once chiral symmetry is broken by a mass term, further non-chiral interactions are generated.
Within the corresponding infinite-dimensional space of couplings, a sorting principle is provided
by the canonical dimensionality of couplings.
Perturbatively, only those couplings with vanishing or positive mass dimensionality can be relevant,
i.e., can survive at low energies.
Quantum-gravity effects could shift perturbatively slightly irrelevant couplings into relevance.
For instance, in asymptotically safe gravity the relevant operators include several dimension-4-operators.
Nonetheless, the departure from canonical scaling appears to be small and the canonical
dimensionality remains a useful guiding principle,
see, e.g., \cite{Falls:2013bv,Falls:2014tra,Narain:2009fy,Narain:2009gb,Eichhorn:2016esv}.
Thus, we analyze a truncated effective dynamics for $N_f$ fermions containing all fermion bilinears
with canonical dimensionality $\leq 5$, i.e., including couplings with dimensionality $\geq -1$
\begin{align}
 \notag
 \Gamma_k = {}& \Gamma_{\rm grav} + i Z_{\psi} \int d^4x\, \sqrt{g}\, \bar{\psi}^i \slashed{\nabla}\psi^i
 + i \bar{m}_{\psi} \int d^4x \sqrt{g} \bar{\psi}^i\psi^i
 \\
 {}& + i \bar{\xi} \int d^4x \sqrt{g} R \bar{\psi}^i\psi^i
 + i \bar{\zeta} \int d^4x \sqrt{g} \bar{\psi}^i \nabla^2 \psi^i.
\end{align}
Herein, $\Gamma_k$ is the scale-dependent effective action that contains the effect of quantum fluctuations
with momenta above $k$, only.
In quantum gravity, the introduction of a scale relies on a dynamically generated
background metric $\bar{g}_{\mu \nu}$.
For the covariant derivative of the fermions we use the spinbase invariant formalism
\cite{Weldon:2000fr,Gies:2013noa}.
The kinetic term features a chiral $U(N_f)_L \times U(N_f)_R$ symmetry, under which left- and
right handed fermions transform separately.
This symmetry is broken explicitly to a global $U(N_f)$ flavor symmetry by the mass term and the
non-minimal interactions.

We employ the functional Renormalization Group \cite{Wetterich:1992yh}
that is well-suited to trace fixed points away from the critical
dimensionality of a given model and discover asymptotic safety, see, e.g., \cite{ASexamples}.
The Wetterich equation governs the momentum-scale dependence of the effective dynamics,
encoded in $\Gamma_k$ \cite{Wetterich:1992yh},
\begin{align}\label{eq:Wetterich}
\partial_t\Gamma_k = \frac{1}{2}\text{STr}\left[(\Gamma_k^{(2)}+R_k)^{-1}\partial_t R_k\right],
\end{align}
with $\partial_t=k\partial_k$, see also \cite{Morris:1993qb, Ellwanger:1993mw}. 
In Eq.~\eqref{eq:Wetterich}, $R_k (\Delta)$ is an IR regulator that provides a
momentum-shell wise integration of the path integral:
IR-modes for which the generalized ``momentum" $\Delta<k^2$ are suppressed.
The supertrace $\rm STr$ implements a summation/integration over the discrete/continuous eigenvalues of
the field-dependent regularized propagator $(\Gamma_k^{(2)}+R_k)^{-1}$.
The $\rm STr$ reduces to a sum over Lorentz and internal indices and a momentum integration
for the case of a flat background, where $\Delta \rightarrow p^2$ (in the absence of gauge fields),
with an additional negative sign for fermionic fields. 
For further details, see \cite{Benedetti:2010nr}, for reviews \cite{Berges:2000ew, Polonyi:2001se,
Pawlowski:2005xe, Gies:2006wv, Delamotte:2007pf, Rosten:2010vm, Braun:2011pp}
and specifically for gravity \cite{ASreviews}, following Reuter's seminal work \cite{Reuter:1996cp}.

\begin{figure}[!t]
  \includegraphics[width=0.2\linewidth]{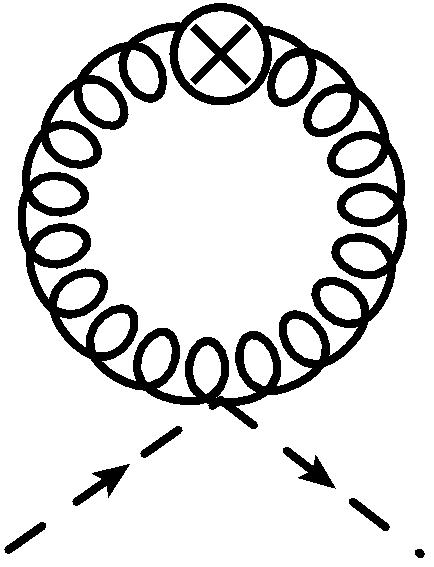}
 \quad \quad
  \includegraphics[width=0.3\linewidth]{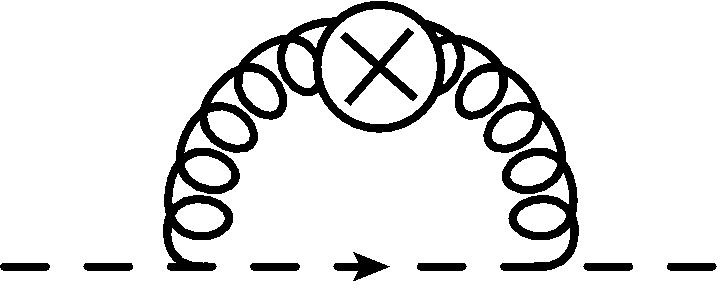}
 \quad\quad
  \includegraphics[width=0.3\linewidth]{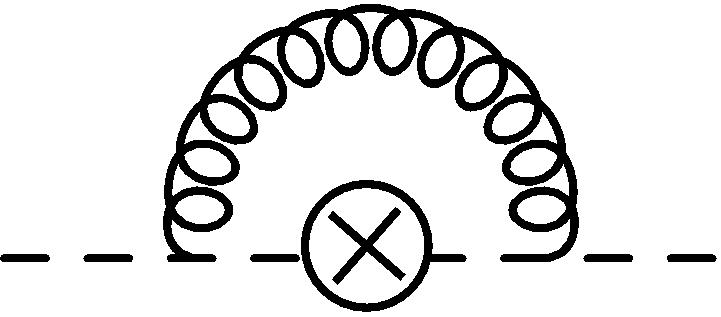}
\caption{\label{fig:flowdiags}
Three diagrams drive the RG flow in the fermionic couplings.
For the two-vertex diagram, the regulator insertion, denoted by a crossed circle,
can be found on either of the internal propagators.
Curly lines denote the metric propagator and dashed lines the fermions.%
}
\end{figure}

To set up the RG flow for gravity, we use the background field method \cite{Abbott:1980hw}
and split the metric into background and fluctuation
\be
g_{\mu \nu} = \bar{g}_{\mu \nu} + \sqrt{16 \pi G}\, h_{\mu\nu}.
\ee
We work with the Einstein-Hilbert action
\be
\Gamma_{\rm grav}=\frac{-1}{16 \pi G}\int d^4x \sqrt{g}\left(R-2 \bar{\lambda} \right) + S_{\rm gf},
\ee
with a gauge fixing term $S_{\rm gf}= \frac{1}{32 \pi \alpha} \int d^4x \sqrt{\bar{g}}
\bar{g}^{\mu \nu}F_{\mu}F_{\nu}$ with $F_{\mu} = \left(\delta_{\mu}^{\lambda}\bar{D}^{\kappa}
- \frac{1}{4}\bar{g}^{\kappa \lambda}\bar{D}_{\mu} \right)h_{\kappa \lambda}$ and gauge parameter $\alpha = 0$.
The Renormalization Group flow for a Litim-type regulator \cite{Litim:2001up}, appropriately chosen
for fermions \cite{Dona:2012am}, in the fermionic sector is driven by two diagrams, cf.~Fig.~\ref{fig:flowdiags}.

We use dimensionless couplings and normalize the kinetic term to its canonical form, obtaining
\begin{eqnarray}
\bar{m}_{\psi} &=& Z_{\psi} m_{\psi} k, \quad\bar{\xi}=Z_{\psi} \frac{\xi}{k},
\quad \bar{\zeta} = Z_{\psi} \frac{\zeta}{k},\nonumber
\\
 G &=& \frac{g}{k^2}, \quad \bar{\lambda} = \lambda\, k^2. 
\end{eqnarray}
\section{Results: Asymptotic safety with heavy and light fermions}\label{sec:ASferm}
In asymptotically safe gravity, a UV completion of the low-energy effective field theory for the metric
is provided by an interacting fixed point of the RG flow, generalizing the powerful concept
of asymptotic freedom to a quantum-gravitational setting.
Previous results indicate that chiral symmetry is not broken spontaneously by asymptotically safe gravity
\cite{Eichhorn:2011pc,Eichhorn:2011ec,Meibohm:2016mkp}.
Here, we go one step further and consider explicit breaking terms. 
We discover two asymptotically safe fixed points.
Both provide a viable generalization of the well-known pure-gravity fixed point \cite{Reuter:1996cp,Falls:2013bv,Falls:2014tra,ASall},
cf.~Tab.~\ref{tab:FPs}.
\begin{table*}
\begin{tabularx}{0.95\textwidth}{*{2}{c|} *{11}{>{\centering\arraybackslash}X|}}
fixed point      & symmetry for fermions & $g_{\ast}$ & $\lambda_{\ast}$ & $m_{\psi\, \ast}$ & $\xi_{\ast}$ & $\zeta_{\ast}$ & $\eta_{\psi}$
& $\theta_1$ & $\theta_2$ & $\theta_3$ & $\theta_4$ & $\theta_5$ \\
\hline\hline
chiral non-Gau\ss{}ian & chiral & 2.52 & -0.42 & 0  & 0    & 0     & -0.17 & 3.54 & 1.34 & 0.84 & -0.73 & -1.27 \\ \hline
chiral non-Gau\ss{}ian & chiral & 2.52 & -0.42 & 0  & --   & --    & -0.17 & 3.54 & 1.34 & 0.84 & --     & --     \\ \hline
chiral non-Gau\ss{}ian & chiral & 2.52 & -0.42 & -- & 0    & --    & -0.17 & 3.54 & 1.34 & --    & -0.69 & --     \\ \hline 
chiral non-Gau\ss{}ian & chiral & 2.52 & -0.42 & -- & --   & 0     & -0.17 & 3.54 & 1.34 & --    & --     & -1.30 \\ \hline
\hline
non-Gau\ss{}ian  & none   & 1.00 &-0.27  & 1.01  & 1.10 & -2.49 & -0.56 & 3.65 & 1.66 & 0.59 & \multicolumn{2}{c|}{-2.50 $\pm$ i 1.60} \\ \hline
non-Gau\ss{}ian  & none   & 2.52 & -0.41 & --     & 0.74 & -- & -0.15 & 3.54 & \multicolumn{2}{c|}{1.37 $\pm$ i 0.04} & -- & -- \\ \hline \hline
\end{tabularx}
\caption{\label{tab:FPs}
We show the fixed-point values and critical exponents at the chiral non-Gau\ss{}ian fixed point,
as well as at the non-Gau\ss{}ian fixed point which explicitly breaks chiral symmetry.
For results in smaller truncations missing couplings and exponents are denoted by a ``--''.
}
\end{table*} 
Towards the IR, the RG flow stays within the critical surface of an UV fixed point,
if it is fine-tuned to that surface in the UV.
This happens automatically, if the flow is set to start at an UV fixed point.
On the other hand, UV-relevant directions in the space of couplings are IR-repulsive.
This will be decisive for the status of chiral symmetry in quantum gravity.
We define the critical exponents as minus the eigenvalues of the stability matrix such
that UV-relevant directions have $\theta>0$, 
\bea
  \theta_I = -{\rm eig} \left(\frac{\partial\beta_{g_i}}{\partial g_j} \right)_{g_n =g_{n}^{\ast}},
\eea
where $g_{i} = (g, \lambda, m_{\psi}, \xi,\zeta)_{i}$.

As expected due to the preservation of global symmetries in the Wetterich equation, one of the fixed points
is governed by an enhanced chiral symmetry and thus enforces a vanishing fermionic mass term at the
microscopic level, reproducing the fixed point discussed in detail in \cite{Dona:2013qba}.
Here, we discover that it features three relevant directions, if we allow non-chiral fluctuations.
Further, $\xi$ and $\zeta$ align quite well with the irrelevant directions.
Thus, both of these couplings are automatically forced to remain zero also in the IR.
On the other hand, the mass operator is UV-relevant.
For the RG flow towards the IR, our results thus indicate that reaching the chirally symmetric
regime requires the tuning of parameters, i.e., chiral symmetry does not act as an infrared attractor.
Thus, chiral symmetry is not an IR emergent phenomenon in asymptotic safety,
which would have guaranteed the existence of light fermions.
However, if chiral symmetry is an exact symmetry of the microscopic theory,
it remains unaffected by gravity fluctuations in the asymptotic-safety scenario.
Hence, asymptotic safety is \textit{compatible} with the existence of light fermions.

The second fixed point is fully interacting and features a nonvanishing fermion mass, cf.~Tab.~\ref{tab:FPs},
generically also resulting in a finite mass in the IR.
It again exhibits three relevant directions.
The third relevant direction at the non-chiral fixed point is a non-trivial superposition of
the fermionic couplings.
In particular, this fixed point cannot be discovered in smaller truncations with $\xi=0$, whereas it
appears in truncations with $m_{\psi}=0$.
Both fixed points conform to our expectation that the canonical dimensionality provides guidance to
estimate whether a coupling is relevant at an interacting fixed point.

Accordingly asymptotic safety does not restrict the nature of fermions,
as there are interacting fixed points with broken and unbroken chiral symmetry.
For the latter, the mass can remain zero on all scales.
Thus, the asymptotic-safety scenario is robust under the inclusion of chiral as well as non-chiral
fermion fluctuations.
\subsection{Several light and heavy fermions}
\begin{figure}[!t]
\quad\quad\includegraphics[width=0.9\linewidth]{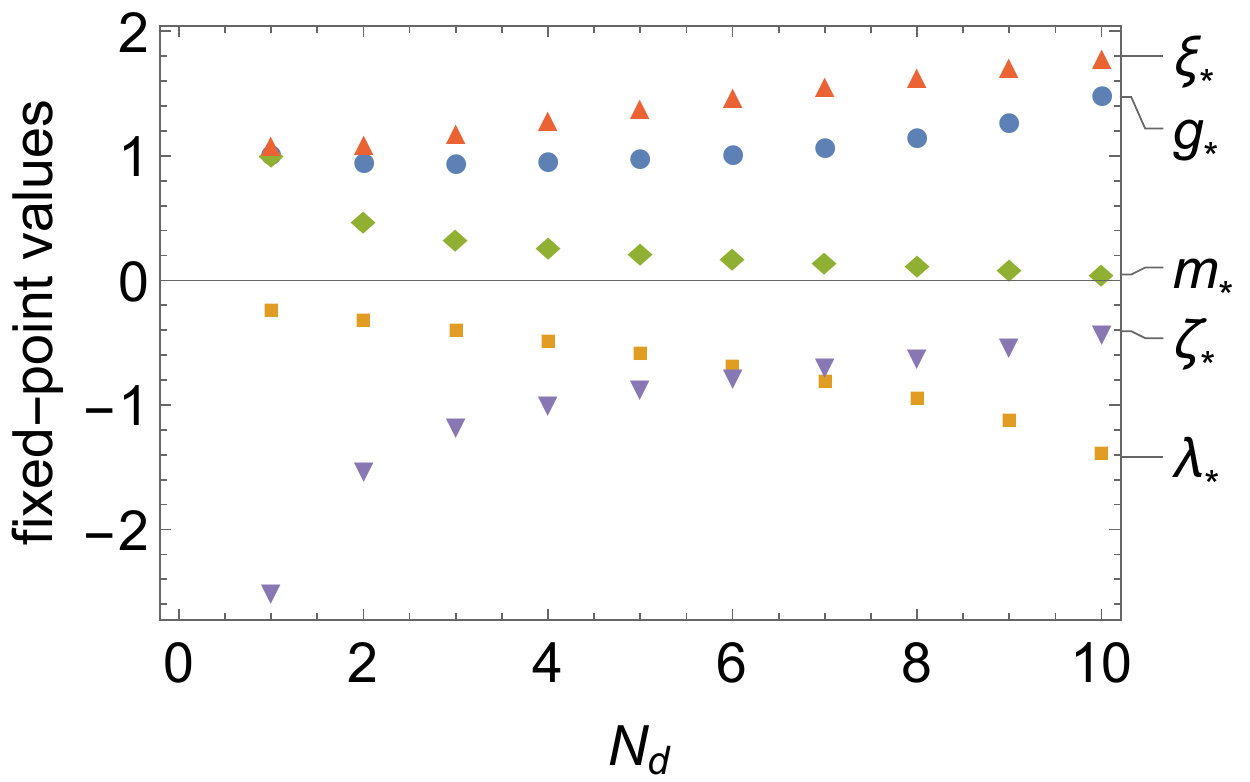}
\\
\includegraphics[width=0.8\linewidth]{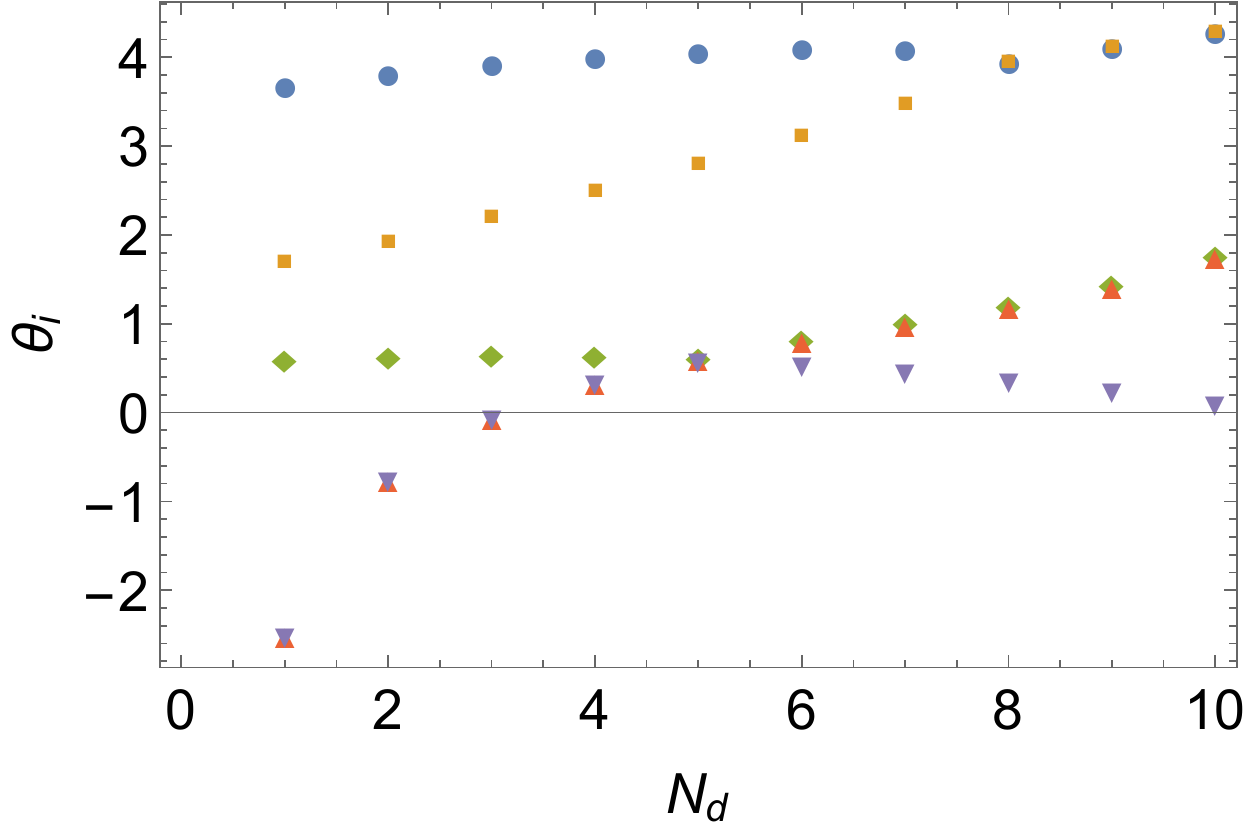}
\caption{\label{fig:fixedpointNd}
We show fixed-point values (upper panel) and critical exponents (lower panel) as a function of $N_d$.%
}
\end{figure}

Here, we focus on the properties of the non-chiral fixed point as a function of $N_d$ non-chiral fermions,
cf~Fig.~\ref{fig:fixedpointNd}, as the chiral fixed point has already been analyzed in
\cite{Dona:2013qba, Dona:2014pla,Meibohm:2015twa}.
The fixed-point values and critical exponents depend on $N_d$, but stay bounded
until the fixed point disappears into the complex plane at $N_d \approx 10$.
There is no obvious indication of a breakdown of our truncation.
Accordingly, we tentatively interpret our results as a hint that nonchiral fermions are only
compatible with asymptotic safety for small fermion numbers.
Already at $N_d \simeq 3$, the nature of the fixed point changes, as it becomes less predictive and
exhibits two further relevant directions. 
Thus, the non-chiral fixed point cannot be continued with real couplings up to $N_d =22.5$,
which is the number of fermions in the standard model.
If this property persists in more extended approximations and under the inclusion of further matter fields,
as in \cite{Dona:2013qba, Dona:2014pla}, this provides a strong link between particle physics and gravity:
If quantum gravity is an asymptotically safe field theory, the fermionic content of the standard model
-- or of any grand unified theory with a large number of fermions -- has to be chiral.
\subsection{UV complete dark matter models}
As we observe two fixed points, one that preserves chiral symmetry and one that does not, an intriguing scenario
is conceivable that could provide a UV completion for fermionic models of dark matter \cite{Kim:2006af,Kim:2008pp}:
Of the $N_{f} + N_{d}$ fermions, a subset of $N_f$ fermions is described by a chirally symmetric fixed point.
On the other hand, $N_d$ fermions feature a nonvanishing microscopic mass and a nonzero nonminimal coupling.
Their mass is not restricted by the symmetry of the microscopic action,
and is a UV-relevant direction of the non-chiral fixed point, and can thus be chosen freely in the IR.
The $N_d$ fermions should be uncharged under the gauge groups of the standard model,
whereas the $N_f$ fermions should have standard-model-like charges.
Accordingly, the $N_d$ massive fermions would interact only weakly (namely only through quantum-gravity
induced interactions, e.g., of the type discussed in \cite{Eichhorn:2016esv}), and might become
candidates for dark matter.
To test the viability of this exciting scenario goes beyond the scope of this paper, so
to exemplify our idea, we explore a toy model with $N_f=1$ and $N_d=1$, which features a fixed point at 
\begin{align}
 \begin{aligned}
    &g^{\ast}_{\phantom{dd}} = 1.36,
    \\
    &\zeta_{d}^{\ast} = -1.81,
    \\
    &\eta_{\psi,d} = -0.38,
 \end{aligned}
\,\,
 \begin{aligned}
    &\lambda^{\ast} = -0.44,
   \\
    &m_{f}^{\ast} = 0,
    \\
    &\eta_{\psi,f} = -0.09.
 \end{aligned}
\,\,
 \begin{aligned}
    &m_{d}^{\ast} = 0.59,
   \\
    &\xi_{f}^{\ast} = 0,
    \\
    &\vphantom{\eta_{\psi,d} = -0.38}
 \end{aligned}
 \,\,
  \begin{aligned}
    &\xi_{d}^{\ast} = 0.97,
   \\
    &\zeta_{f}^{\ast} = 0,
    \\
    &\vphantom{\eta_{\psi,d} = -0.38}
 \end{aligned}
\end{align}
The set of critical exponents features just four relevant ones, corresponding exactly to the number of
relevant operators in the system ($\sqrt{g} R,\, \sqrt{g},\, \sqrt{g} \bar{\psi}_f\psi_f,
\, \sqrt{g} \bar{\psi}_d\psi_d$):
\begin{align}
 \begin{aligned}
   &\theta_1=3.74,
  \\
   &\theta_5=-0.86,
 \end{aligned}
\quad
 \begin{aligned}
   &\theta_2=1.66,
  \\
   &\theta_6= -1.14,
 \end{aligned}
\quad
 \begin{aligned}
   &\theta_3 =0.92, \quad \theta_4=0.46,
  \\
   &\theta_{7,8} = -1.20 \pm i\, 1.92.
 \end{aligned}
\end{align}
Although all the non-chiral couplings are set to zero for the light fermion, its fluctuations affect the
running of $g$ and $\lambda$, and thereby shifts the fixed-point values for the complete system,
including those for the non-chiral couplings of the dark fermion.
Whether such a model can reproduce the correct relic density of dark matter fermions, and is consistent 
with all current exclusion limits on fermionic dark matter \cite{Beniwal:2015sdl}, is an important question,
but beyond the scope of the present work.
\section{Consequences for quantum gravity}\label{sec:QGgeneral}
Broadening the scope of our work, we consider models of quantum gravity beyond asymptotic safety.
Typically, these models are not formulated in terms of a local continuum quantum field theory for the metric,
as we employ here.
However, any viable model of quantum gravity must reduce to Einstein gravity in the low-energy regime.
Thus, there is a scale $k_{\rm trans}$ at which one can translate from the microscopic model to an
effective description in terms of a local model of metric fluctuations, and arrive at the low-energy
effective field theory for quantum gravity, see, e.g., \cite{EFT}.
Intuitively, this is analogous to the translation between, e.g., a microscopic condensed matter model of,
e.g., electronic interactions, and the effective description in terms of emergent low-energy
degrees of freedom, e.g., phonons.
The microscopic model then provides the values of all couplings in the effective description at the
scale $k_{\rm trans}$.
In the RG language, this determines a starting point in theory space at $k_{\rm trans}$ for the RG flow
towards the IR.
However, from a generic point in theory space, we cannot reach the subspace in which
chiral symmetry would be restored.
This follows, as the stability matrix for that case, $- \partial \beta_{g_i}/ \partial g_j$,
$\vec{g}= ( m_{\psi}, \xi, \zeta )$ features at least one positive, i.e.,
IR-repulsive eigenvalue for all values of $g \in (0,30)$ and $\lambda \in(-2 , 0.4)$.
Accordingly, any microscopic model for which the fermion mass and further chiral-symmetry breaking
effects do not vanish exactly at $k_{\rm trans}$ will generically feature rather large
fermion masses of the order of $k_{\rm trans}$ in the IR.
Thus, a quantum gravity model that cannot enforce chiral symmetry at the microscopic level will not
exhibit emergent chiral symmetry in the infrared.
In other words, these models will not be compatible with the existence of light fermions as those
in the standard model.
Our result highlights that once lost, chiral symmetry is not an emergent symmetry of the RG flow
to the infrared in the effective theory regime.
In fact, this could present a severe restriction of quantum gravity models based on a discrete spacetime,
see, e.g., \cite{Gambini:2015nra,Barnett:2015ara}:
Under certain assumptions, chiral fermions do not exist on a regular
lattice \cite{Nielsen:1981hk,Nielsen:1980rz,Nielsen:1981xu}.
To circumvent this, the lattice Dirac operator should satisfy the Ginsparg-Wilson
relation \cite{Ginsparg:1981bj}, implying the existence of a continuous symmetry \cite{Luscher:1998pqa}.
Thus, chiral symmetry does not emerge in the continuum automatically,
but requires the microscopic dynamics to obey a discrete version of this symmetry.
Our results imply that quantum fluctuations of the spacetime do not remove the need
for fine-tuning in order for chiral symmetry to emerge.
This might in the future be tested in explicit Monte Carlo simulations of discrete quantum
gravity \cite{Ambjorn:1998xu,Ambjorn:2000dv,Ambjorn:2001cv,Ambjorn:2012jv}.

Thus the observational fact that light fermions exist can be translated into a requirement
for any microscopic model of quantum gravity:
Whatever the exact form of the dynamics, chiral symmetry must be built into the model for
those degrees of freedom that end up being Standard Model fermions at low energies.
\section{Conclusions}
We highlight how the low-energy properties of matter can impose nontrivial observational constraints
on quantum gravity models.
In particular, we exploit the existence of light fermions, that is tied to chiral symmetry,
and study the interplay of chiral symmetry with quantum gravity.
We discover that chiral symmetry does not act as an infrared attractor, i.e.,
quantum gravity fluctuations prevent the emergence of chiral symmetry in the RG flow towards the infrared.
Thus, viable models of quantum gravity must have chiral symmetry built in at the microscopic level.
As a specific example, asymptotically safe gravity features two UV fixed points,
which provide UV completions for quantum gravity:
One of them features chiral symmetry, and is compatible with a vanishing fermion mass on all scales.
The second fixed point, with broken chiral symmetry, disappears into the complex plane for more
than ten fermion species.
Thus, the chiral nature of fermions in the Standard Model might be enforced by asymptotic safety.
Finally, we also discover fixed points with $N_f$ chiral and $N_d$ massive fermions, which might provide
a template for an asymptotically safe model of Standard Model matter and massive, fermionic dark matter.

\emph{Acknowledgements}
We thank H.~Gies for discussions on this topic and comments on the draft.
This research was supported by the Perimeter Institute for Theoretical Physics during its initial stages.
S.~L.~thanks the Perimeter Institute for Theoretical Physics as well as Imperial College for hospitality.
S.~L.~acknowledges support by the DFG under Grants No.~GRK1523/2, Gi328/7-1.
A.~E.~acknowledges support by an Imperial College Junior Research Fellowship during a part of this work,
and support by the DFG through the Emmy-Noether program under grant no. Ei-1037-1.

\end{document}